\documentclass[aps,prd,twocolumn,amsmath,amssymb,nofootinbib]{revtex4-2}
\usepackage{bm,siunitx,booktabs,url,hyperref,graphicx}
\hypersetup{colorlinks=true,linkcolor=blue,citecolor=blue,urlcolor=blue}
\newcommand{\lkk}{\left[}
\newcommand{\rkk}{\right]}

\newcommand{\beq}{\begin{equation}}
\newcommand{\beqa}{\begin{eqnarray}}
		  \newcommand{\eeq}{\end{equation}}
\newcommand{\eeqa}{\end{eqnarray}}

\begin{document}

\title{Probing the Orientation Distribution of Nearby Double White Dwarfs through Gravitational Waves with LISA}
\author{N. Seto$^1$ and G.C. Liu$^2$}
\affiliation{$^1$Department of Physics, Kyoto University, 
Kyoto 606-8502, Japan\\
$^2$Department of Physics, Tamkang University, Tamsui, New Taipei City 25137, Taiwan
}
\date{\today}

\begin{abstract}
We present a framework to study the orientation distribution of
nearby double white dwarf binaries with LISA, explicitly accounting for
the fourfold degeneracy in the polarization angle. 
Exploiting the belt-like sky distribution of the nearby sample along
the Galactic plane, we
show that the low-multipole extraction can be recast as an effectively
one-dimensional Fourier analysis along Galactic longitude, within a
spherical-harmonic expansion based on the Wigner-\(D\) formalism. The
symmetry with respect to the Galactic plane governs the sensitivity:
modes even under reflection across the plane remain accessible, while
odd modes are strongly suppressed. For a fiducial sample of about 500
binaries within \(\lesssim 4\) kpc, corresponding to roughly 5\% of the
resolved Galactic population, the method yields an uncertainty of order
0.02 for the even coefficients with \(\ell \leq 3\), and is readily
applicable to the forthcoming LISA catalog.
\end{abstract}

\maketitle

\section{Introduction}
The orientation distribution of the orbital angular momenta of binary
systems carries valuable information about their formation and
evolutionary processes. Observational studies of Galactic binary
orientations have been conducted for more than a century (see
\cite{1929AJ.....40...11C} for early studies). More recently,
Agati et al.~\cite{2015A&A...574A...6A} analyzed 95 binaries within
18 pc of the Sun and found their unit angular momentum vectors to be
consistent with a random distribution (see also \cite{arenou2023gaia}
for potential selection effects in the Gaia survey). In contrast,
Tan et al.~\cite{2023ApJ...951L..44T} examined 14 planetary nebulae near
the Galactic bulge that are thought to host short-period binaries.
Planetary nebulae consist of gas expelled during the formation of white
dwarfs, and the symmetry axes of the nebulae are expected to align with
the orbital angular momenta of the central binaries. Tan et
al.~\cite{2023ApJ...951L..44T} reported that these axes tend to be
parallel to the Galactic plane at the \(5\sigma\) level. These
contrasting results raise the possibility that binary orientations
depend on environment, possibly reflecting effects such as magnetic
fields present at formation~\cite{2023ApJ...951L..44T}. 
A comparison between local and bulge populations may therefore provide a
useful test of whether orientation distributions are universal or depend
on their formation environment.

Nevertheless, existing observational studies are limited by small sample
sizes, leaving significant room for further development. It is therefore
important to carry out statistical analyses across different Galactic
environments and binary populations, using observational probes
complementary to traditional electromagnetic methods.

A key advantage of gravitational-wave (GW) observations is their
geometric sensitivity to the orbital configuration of binaries (see
e.g., \cite{2014grav.book.....P}). {For a sufficiently well measured
circular binary, the GW signal constrains the inclination angle \(I\),
measured with respect to the line of sight, and the polarization angle
\(\psi\), which specifies the azimuthal direction of the projected
angular momentum on the sky. At the quadrupole-waveform level, the
polarization angle has a fourfold degeneracy,
\(\psi\to\psi+k\pi/2\) with \(k=0,1,2,3\)
\cite{Cornish:2003vj}.} Thus GW observations provide a
fundamentally different, geometric probe of binary orientations,
{while also imposing a characteristic projection problem that must
be treated statistically.}

The planned space-based GW observatory LISA is expected to detect
\(\sim10^4\) Galactic double white dwarfs (DWDs) above \(\sim3\,\mathrm{mHz}\),
providing an essentially complete survey of this Galactic population in
this band
\cite{LISA:2022yao,Nissanke:2012eh,Korol:2017qcx,Lamberts:2019nyk}
(see also \cite{Hu:2017mde,TianQin:2015yph,Huang:2020rjf} for Taiji
and TianQin). LISA can infer distances to inspiraling DWDs from their
amplitudes and orbital decay rates with useful precision
\cite{Takahashi:2002ky}, and can determine sky locations primarily
through Doppler modulations \cite{Cutler:1997ta}. {The resulting
resolved DWD catalog will therefore provide a large statistical sample
for studies of compact-binary orientations.}

Motivated by Tan et al.~\cite{2023ApJ...951L..44T},
Seto~\cite{Seto:2024odc} developed a spherical-harmonic formalism to
probe the quadrupolar pattern in the orientation distribution of bulge
DWDs with LISA, under the assumption of rotational symmetry (see also
\cite{Seto:2024wrw} for dipole contributions). While those studies
\cite{Seto:2024odc,Seto:2024wrw} focused on the bulge, LISA will also
resolve a large number of nearby disk DWDs. In view of the contrast
between the results of \cite{2015A&A...574A...6A} for local
main-sequence binaries and \cite{2023ApJ...951L..44T} for bulge
planetary nebulae, which are evolutionarily related to compact white
dwarf binaries, the nearby disk DWD population represents a natural and
compelling target for orientation studies.

{The physical quantity of interest is not the distribution of the
source-frame angles \(I\) and \(\psi\) itself, but the distribution of
the orbital angular momentum direction \(\hat{\bm l}\) in Galactic
coordinates. In this paper we denote by
\(Q_{\rm loc}(\hat{\bm l})\) the orientation distribution averaged over
the selected local volume.}

{For an old disk population, Galactic orbital phase mixing motivates
a fiducial scenario in which non-axisymmetric components around the
Galactic rotation axis are suppressed. The axisymmetric coefficients
\(a_{\ell 0}\) are therefore the primary targets. As discussed below, however, the dipole coefficient \(a_{10}\) is
strongly suppressed by the nearly midplane-symmetric projection of the
nearby disk sample considered here. The
quadrupolar coefficient \(a_{20}\) is therefore the most promising
axisymmetric target in the present setup. It directly describes the type
of alignment suggested by the planetary-nebula result: \(a_{20}>0\) corresponds to an excess of orbital angular momentum
vectors toward the Galactic poles, whereas \(a_{20}<0\) corresponds to
a preference for orbital angular momentum vectors lying in the Galactic
plane. We nevertheless keep the general \(m\)
dependence in the formalism below, because it clarifies how the
observational projection selects, suppresses, or mixes the different
components.}

In this paper we assess the prospects for measuring
\(Q_{\rm loc}(\hat{\bm l})\) with nearby
(\(\lesssim4\)\,kpc) DWDs detected by LISA above \(3\)\,mHz. The
expected size of this local sample is \(M_{\mathrm{tot}}\sim500\),
corresponding to about 5\% of the entire resolved Galactic DWD
population (estimated by integrating the Galactic model in
\cite{Nissanke:2012eh}). Because the survey region is small compared with the Galactic scale, the
extracted coefficients should be interpreted as sample-averaged local
moments. On the sky, however,
these systems form a thin belt along the Galactic disk that encircles
the celestial sphere, in sharp contrast to the bulge DWDs, which are
strongly concentrated in a restricted region. This geometrical
difference changes the low-multipole extraction problem: for the
nearby disk sample, it can be recast as an effectively one-dimensional
Fourier analysis along Galactic longitude.

The structure of this paper is as follows. Section~II introduces the
source-frame orientation angles and the local Galactic-frame distribution
\(Q_{\rm loc}(\hat{\bm l})\), and relates them using a
Wigner-\(D\) rotation formalism while treating the fourfold degeneracy
of the polarization angle. Section~III shows how the belt-like sky
distribution of nearby disk DWDs reduces the low-multipole extraction
to a one-dimensional Fourier problem along Galactic longitude.
Section~IV discusses shot noise due to the finiteness of the local DWD
sample. Section V discusses limitations and possible extensions. Section~VI
provides a short summary.
% notation
% \newcommand{\hl}{\hat{\bm l}}
% \newcommand{\hn}{\hat{\bm n}}

% Sun-centered local volume and position-independence
\section{Distribution functions}

{This section sets up the basic geometrical notation and projection
formalism on which the later disk-limited analysis is built. The target
quantity is the local distribution of orbital angular momentum
directions in Galactic coordinates, whereas the orientation information
obtained for each binary is naturally described by the source-frame
angles \(I\) and \(\psi\). Our task is to relate these two descriptions
and to identify how the fourfold degeneracy of the polarization angle
restricts the observable components. The Wigner-\(D\) notation
introduced below is only a compact way of expressing the rotation
between the Galactic and source frames. In Sec.~\ref{sec:gal}, this
general projection formalism will be specialized to the belt-like sky
distribution of nearby disk DWDs.}

\subsection{Conventions and basic definitions}\label{sec:conv}
We first summarize the basic notation for the sky positions and orbital
orientations of DWDs.

The sky direction is denoted as $\hat{\bm n}=(b,\varphi)$ with Galactic latitude $b$
and longitude $\varphi$ ($\varphi=0$ pointing toward the Galactic center).
In relation to the conventional definition of spherical harmonics,
we introduce the polar angle
\[
\theta=\frac{\pi}{2}-b,
\]
so that $\theta=0$ corresponds to the North Galactic pole.
This definition ensures that our coordinate system matches the standard physics
convention, where $Y_{\ell m}(\theta,\varphi)$ are defined with $\theta$ as the
polar angle from the $z$-axis. The $x$-axis is directed toward the Galactic center. 

We denote by $\hat{\bm l}$ the unit vector along the orbital angular momentum.
In general, the orientation distribution may depend on the position of the
binary. We denote by
\begin{equation}
Q(\hat{\bm l}:\hat{\bm n},d)
\end{equation}
the distribution function of $\hat{\bm l}$ for DWDs at heliocentric
distance $d$ and sky direction $\hat{\bm n}$.

In this work we select DWDs within $d_{\max}=4\,\mathrm{kpc}$.
This distance is smaller than the Galactocentric distance
$R_0\simeq8.3\,\mathrm{kpc}$, and the survey volume subtends a half-angle
of only $\arcsin(d_{\max}/R_0)\sim30^\circ$ as seen from the Galactic center.
{We do not attempt to resolve possible spatial variations of the
orientation distribution within this volume. Instead, we define
\(Q_{\rm loc}(\hat{\bm l})\) as the effective local distribution averaged
over the selected sample. The harmonic coefficients introduced below
should therefore be interpreted as sample-averaged local moments of the
orientation distribution.}

Next, we expand this local distribution in spherical harmonics defined
with respect to the Galactic coordinate system introduced above:
\begin{equation}
Q_{\rm loc}(\hat{\bm l})=\sum_{\ell m} a_{\ell m}\,Y_{\ell m}(\hat{\bm l}).
\label{eq:Qgal}
\end{equation}
{As discussed in the Introduction, the axisymmetric coefficients
\(a_{\ell0}\) are natural physical targets for a phase-mixed old disk
population. We nevertheless keep the general \(m\) dependence in
Eq.~(\ref{eq:Qgal}) in order to describe the observational projection
without imposing axisymmetry.}
In the following sections we discuss how these coefficients $a_{\ell m}$
can be determined observationally.

\subsection{Source frame and polarization basis}\label{sec:source}

{The coefficients \(a_{\ell m}\) in Eq.~(\ref{eq:Qgal}) are defined
in Galactic coordinates. By contrast,} GW observables for each binary
are most naturally described in the source frame, whose polar axis is
taken along the line of sight \(+\hat{\bm n}\). {We denote by \(I\)
the inclination angle between \(\hat{\bm l}\) and \(\hat{\bm n}\),
defined by}
\begin{equation}
\cos I \equiv \hat{\bm l}\!\cdot\!\hat{\bm n},
\end{equation}
so that \(I=0\) (\(\hat{\bm l}=\hat{\bm n}\)) and \(I=\pi\)
(\(\hat{\bm l}=-\hat{\bm n}\)) are face-on, while
\(I=\pi/2\) is edge-on. Note that some GW analyses adopt the opposite
convention \(\cos I \equiv -\,\hat{\bm l}\!\cdot\!\hat{\bm n}\).

The polarization angle \(\psi\) {specifies the azimuthal direction
of the projection of \(\hat{\bm l}\) on the plane perpendicular to
\(\hat{\bm n}\). We measure \(\psi\)} from the local meridional basis
\(\mathbf e_\theta\). With this choice the Wigner-\(D\) Euler angles can
be taken as \((\varphi,\theta,0)\) in the \(zyz\) sequence \cite{Edmonds1957}, keeping
later expressions compact.

The local distribution \(Q_{\rm loc}\), expanded in Galactic
coordinates, can then be expressed in the source basis
\(Y_{\ell s}(I,\psi)\) via the Euler angles \((\varphi,\theta,0)\) as
\begin{equation}
Q_{\rm loc}(I,\psi:\hat{\bm n})=\sum_{\ell m s}
a_{\ell m}\,E^\ell_{ms}(\hat{\bm n})\,Y_{\ell s}(I,\psi),
\label{q0}
\end{equation}
where
\begin{equation}
E^\ell_{ms}(\hat{\bm n})=D^\ell_{ms}(\varphi,\theta,0)
=\sqrt{\frac{4\pi}{2\ell+1}}\,{}_{-s}Y_{\ell m}^*(\hat{\bm n}) .
\label{eqq6}
\end{equation}
{The first equality in Eq.~(\ref{eqq6}) is the standard
Wigner-\(D\) rotation formula familiar from angular-momentum theory
\cite{Edmonds1957}. The second equality uses the conventional relation
between Wigner-\(D\) matrices and spin-weighted spherical harmonics
\cite{Thorne:1980rev}.}
The coefficients \(E^\ell_{ms}\) play the role of geometrical kernels,
describing how the Galactic-frame azimuthal modes \(m\) are mapped into
the source-frame azimuthal modes \(s\), with \(|s|\le \ell\).
{The spin-weighted spherical harmonics in Eq.~(\ref{eqq6}) are used
here only as a compact notation for the Wigner-\(D\) rotation
coefficients; no differential or field-theoretic properties of
spin-weighted fields are needed in the present analysis.}

For \(s=0\) the kernel reduces to the usual spherical harmonics,
\begin{equation}
E^\ell_{m0}(\hat{\bm n})=\sqrt{\frac{4\pi}{2\ell+1}}\,
Y_{\ell m}^*(\hat{\bm n}).
\label{EY}
\end{equation}
This special case will be frequently used in the following analysis.

\subsection{Fourfold degeneracy}
\label{sec:fold}

For circular binaries in the Newtonian approximation, the GW signal
possesses a fourfold degeneracy under rotations of the polarization
angle \(\psi\). The waveforms with the following four polarization angles
\[
\psi,\ \psi+\frac{\pi}{2},\ \psi+\pi,\ \psi+\frac{3\pi}{2}
\]
are observationally indistinguishable in the quadrupole GW signal
\cite{Cornish:2003vj} (see also \cite{Seto:2025vud,Seto:2025vfg} for
reducing this degeneracy through multimessenger observations and
higher-harmonic GW signals).

To handle this degeneracy, we redefine the distribution function by
folding it with a period \(\pi/2\):
\begin{equation}
\bar Q_{\rm loc}(I,\psi:\hat{\bm n})\equiv
\frac{1}{4}\sum_{k=0}^3
Q_{\rm loc}(I,\psi+k\pi/2:\hat{\bm n}).
\label{eq:Qfold_def}
\end{equation}
The folded distribution \(\bar Q_{\rm loc}\) is kept as a function on
the full interval \(\psi\in[0,2\pi)\), as in the original definition
\(Q_{\rm loc}(I,\psi:\hat{\bm n})\).

{This folding is a projection onto the part of the distribution that
is distinguishable with quadrupole GW data alone. It does not impose an
additional physical symmetry on \(Q_{\rm loc}\), but only averages over
the four polarization-angle representatives that correspond to the same
observed waveform. The discarded components are therefore not recoverable from quadrupole
GW data alone without additional information, such as multimessenger
phase information or higher-harmonic GW signals.}

From the viewpoint of spherical-harmonic expansion, this folding
operation selects only the modes \(s\) satisfying
\[
s \equiv 0 \pmod{4}.
\]
Indeed, from Eq.~\eqref{q0} and \(Y_{\ell s}(I,\psi)\propto e^{i s\psi}\),
\begin{equation}
\sum_{k=0}^3Y_{\ell s}(I,\psi+k\pi/2) \propto e^{i s\psi}
\big(1+e^{i s\pi/2}+e^{i s\pi}+e^{3i s\pi/2}\big),
\end{equation}
which vanishes except for \(s\) divisible by 4. Consequently, the folded
distribution can be written as
\begin{equation}
\bar Q_{\rm loc}(I,\psi:\hat{\bm n})=
\sum_{\ell m}\ \sum_{s=0,\pm4,\pm8,\cdots}
a_{\ell m}\,E^\ell_{ms}(\hat{\bm n})\,Y_{\ell s}(I,\psi),
\label{eq:map2}
\end{equation}
with the kernel \(E^\ell_{ms}(\hat{\bm n})\) defined in
Sec.~\ref{sec:source}.  Since spherical harmonics satisfy \(|s|\leq \ell\), this selection rule
immediately shows that for \(\ell\leq 3\) only the \(s=0\) modes survive,
while at \(\ell=4\) the additional modes \(s=\pm4\) first appear.

{It is useful to separate the coordinate rotation from the information
loss caused by the fourfold degeneracy. Equation~\eqref{q0} represents
a rotation from the Galactic frame to the source frame at each sky
direction \(\hat{\bm n}\). Without the fourfold degeneracy, the full
set of source-frame indices \(-\ell\le s\le \ell\) would be available.
For a fixed \(\hat{\bm n}\), this transformation is just a change of
basis within the same \(\ell\) subspace, and could in principle be
inverted to recover the Galactic-frame coefficients \(a_{\ell m}\).}

{After the folding operation, only the indices
\(s=0,\pm4,\cdots\) remain accessible. Thus the projection over the
source-frame angles \((I,\psi)\) no longer provides enough information,
at a single sky direction, to determine the Galactic-frame coefficients
\(a_{\ell m}\). The remaining information is carried by the dependence
of the kernels \(E^\ell_{ms}(\hat{\bm n})\) on the sky direction. We
make this two-step structure explicit in the next subsection. We then
discuss the ideal full-sky reference case before specializing to the
disk-limited geometry.}

\subsection{Patch-level observables and mode extraction}
\label{sec:patch}

{Equation~\eqref{eq:map2} has a simple inversion structure. The
dependence on the source-frame angles \((I,\psi)\) is used to isolate
the indices \((\ell,s)\), while the dependence on the sky direction
\(\hat{\bm n}\) is used to separate the Galactic-frame index \(m\). We
spell out this two-step structure before specifying the sky coverage of
the sample.}

{Consider a small sky patch \(p\) centered on the direction
\(\hat{\bm n}_p\). The patch is assumed to be smaller than the angular
scale of the modes under consideration, so that the geometrical kernels
\(E^\ell_{ms}(\hat{\bm n})\) can be treated as approximately constant
within the patch. Suppose that \(M_p\) binaries fall in this patch, and
label them by \(i=1,\ldots,M_p\). The harmonic moment of the
source-frame distribution can then be estimated by replacing the
ensemble average over \(\bar Q_{\rm loc}(I,\psi:\hat{\bm n}_p)\) with
the sample average over these binaries. We define}
\begin{equation}
S^\ell_s(\hat{\bm n}_p)
\equiv
\frac{1}{M_p}
\sum_{i=1}^{M_p}
Y^*_{\ell s}(I_i,\psi_i).
\label{eq:patch_estimator}
\end{equation}
{At the level of the harmonic projection, this estimator can be
introduced for any \(s\) with \(|s|\leq \ell\). In the folded
distribution, however, only the components with
\(s=0,\pm4,\pm8,\ldots\) have nonzero expectation values.}

{Using Eq.~\eqref{eq:map2} and treating
\(E^\ell_{ms}(\hat{\bm n})\) as constant inside the patch, we obtain }
\begin{equation}
\begin{aligned}
\mathbb{E}\big[S^\ell_s(\hat{\bm n}_p)\big]
&\simeq
\int
\bar Q_{\rm loc}(I,\psi:\hat{\bm n}_p)
Y^*_{\ell s}(I,\psi)\,
d\cos I\,d\psi  \\
&=
\sum_m a_{\ell m} E^\ell_{ms}(\hat{\bm n}_p).
\end{aligned}
\label{eq:patch_expectation}
\end{equation}
{Here \(\mathbb{E}\) denotes the statistical expectation value.}
{Thus the projection over \((I,\psi)\) isolates the pair
\((\ell,s)\), but leaves a linear combination of the Galactic-frame
coefficients \(a_{\ell m}\). The remaining task is to use the sky
dependence of \(E^\ell_{ms}(\hat{\bm n}_p)\) to separate the
\(m\)-modes. This second step depends on the available sky coverage. If
the sources covered the full sky, the separation would be achieved by
full-sky orthogonality. For the nearby disk sample, however, the
available directions are restricted to a belt around the Galactic plane,
and the corresponding inversion must be adapted to that geometry.}

\subsection{Ideal full-sky deconvolution as a reference case}
\label{sec:seg}

{We first consider the ideal case in which the patch-level quantities
\(S^\ell_s(\hat{\bm n}_p)\) introduced in
Eq.~\eqref{eq:patch_estimator} are available over the full sky with
sufficiently fine patches. This case is not intended to model the
nearby DWD sample considered in this paper. Rather, it provides a useful
reference for the second step of the inversion, namely the separation of
the Galactic-frame index \(m\). For notational simplicity, we write the
resulting continuum limit by replacing the discrete sum over patches
with an integral over \(\hat{\bm n}\).}

{From Eq.~\eqref{eq:patch_expectation} and the relation}
\begin{equation}
E^\ell_{ms}(\hat{\bm n})
=\sqrt{\frac{4\pi}{2\ell+1}}\,
{}_{-s}Y_{\ell m}^*(\hat{\bm n}), \label{ee}
\end{equation}
{the \(m\)-modes can be separated by the full-sky orthogonality of the
spin-weighted spherical harmonics,}
\begin{equation}
\int d\hat{\bm n}\ {}_{-s}Y_{\ell m}(\hat{\bm n})\,
{}_{-s}Y_{\ell' m'}^*(\hat{\bm n})
=\delta_{mm'}\delta_{\ell \ell'} .
\label{nn15}
\end{equation}

{In this idealized limit, we can introduce the sky-aggregated
estimator \(A_{\ell m}\) by combining the continuum version of the
patch-level estimator \(S^\ell_s(\hat{\bm n})\) as}
\begin{equation}
A_{\ell m}
=
\sqrt{\frac{2\ell+1}{4\pi}}
\int
S^\ell_s(\hat{\bm n}){}_{-s}Y_{\ell m}(\hat{\bm n})\,
d\hat{\bm n} .
\label{ort}
\end{equation}
{This definition assumes full-sky coverage with sufficiently fine sky
patches. From Eqs.~\eqref{eq:patch_expectation}, \eqref{ee}, and
\eqref{nn15}, we obtain}
\begin{equation}
\mathbb{E}\big[A_{\ell m}\big]  =a_{\ell m}.
\end{equation}
{Thus, even after the fourfold polarization degeneracy is imposed,
the coefficients \(a_{\ell m}\) would be recoverable in principle if the source directions \(\hat{\bm n}\) were available over the full sky.  In that
ideal limit, the remaining sky dependence of
\(E^\ell_{ms}(\hat{\bm n})\) is sufficient to separate the
Galactic-frame index \(m\).}

{The nearby DWD sample considered in this paper is far from this ideal
full-sky configuration. It is concentrated in a narrow belt around the
Galactic plane, and the estimator \(A_{\ell m}\) cannot be constructed
by a simple full-sky orthogonality relation. In the next section we
therefore replace the full-sky inversion by a disk-limited Fourier
extraction adapted to the belt-like sky distribution.}
The finite number of binaries introduces an additional statistical
uncertainty, which will be discussed in Sec.~\ref{sec:shot}.

\section{Disk-limited Galactic binaries}
\label{sec:gal}

We now specialize the general mode-extraction framework to the nearby
DWD sample. As emphasized above, this sample does not provide full-sky
coverage, but is concentrated in a narrow belt around the Galactic
plane and is essentially absent in the polar directions.

This belt-like geometry changes the second step of the inversion:
although full-sky information is unavailable, the sample still extends
along Galactic longitude. For \(\ell\le3\), the fourfold polarization
degeneracy leaves only \(s=0\). In the thin-disk limit, the remaining
projection kernel has a simple longitude dependence, \(e^{-im\varphi}\).
The recovery of \(a_{\ell m}\) therefore reduces to a one-dimensional
Fourier problem along Galactic longitude, rather than a full-sky
harmonic deconvolution.

In the subsections below, we first introduce a simple disk model for the
spatial distribution of the DWD sample. We then describe the projection
kernels in this disk-limited setup and present the Fourier-based
extraction of the coefficients \(a_{\ell m}\).
\subsection{Galactic model and sample distribution}
\label{sec:model}

As already noted, our analysis is limited to DWDs within
\(d_{\max}=4\,\mathrm{kpc}\), where the disk component dominates over
bulge or halo populations. The vertical scale height of the disk,
\(h_z\simeq0.3\,\mathrm{kpc}\) \cite{Korol:2017qcx}, gives a
characteristic angular thickness
\(b_t\sim h_z/d_{\max}\sim5^\circ\simeq0.09\) at the outer edge of the
sample. This angular scale corresponds to multipoles of order
\(\ell\sim b_t^{-1}\sim10\). Therefore, for the low multipoles
considered here, \(\ell\leq3\), the disk can be treated as thin to
leading order. This estimate is used only to set the scale of the
disk-limited approximation, not to replace the actual latitude
distribution of the sources. In the following, we use the thin-disk
limit to identify the leading projection kernels, and treat the
residual latitude dependence as a subleading correction.

\subsection{Projection kernels in the disk-limited regime}
\label{sec:kernels}

We restrict our discussion to lower multipoles with \(\ell \le 3\).
As explained in Sec.~\ref{sec:fold}, the fourfold degeneracy leaves
only the \(s=0\) components in this range. Using the standard
spherical-harmonic conventions \cite{Edmonds1957}, the relevant
projection kernels are
\begin{equation}
E^{\ell}_{m0}(b,\varphi)\equiv
\sqrt{\frac{4\pi}{2\ell+1}}\,
Y_{\ell m}^\ast\!\left(\tfrac{\pi}{2}-b,\varphi\right).
\label{eq:E_def}
\end{equation}
Using \(Y_{\ell m}(\theta,\varphi)=Y_{\ell m}(\theta,0)e^{im\varphi}\),
the azimuthal dependence factorizes as
\begin{equation}
E^{\ell}_{m0}(b,\varphi)=e^{-im\varphi}\,E^{\ell}_{m0}(b,0).
\label{eq:phi_fact}
\end{equation}

\begin{figure}[t]
  \centering
  \includegraphics[width=0.92\linewidth]{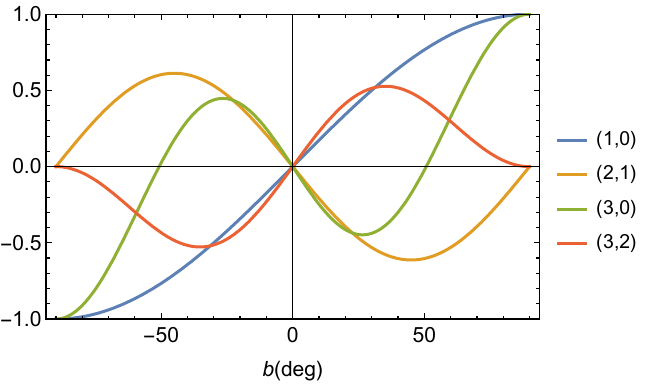}
  \includegraphics[width=0.92\linewidth]{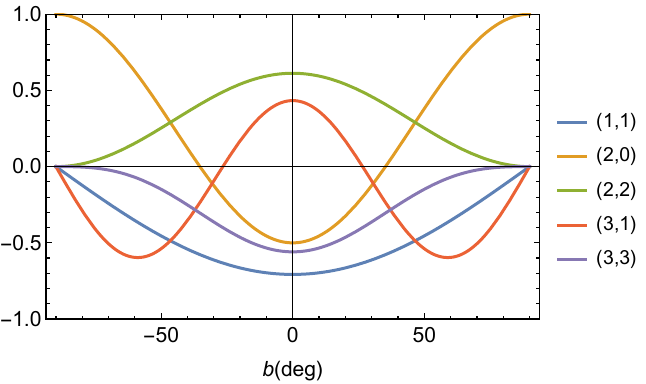}
  \caption{Profiles of the projection kernels
\(E^{\ell}_{m0}(b,\varphi=0)\) as functions of Galactic latitude \(b\)
(in degrees). Upper panel: odd \((\ell+m)\) modes,
\((\ell,m)=(1,0),(2,1),(3,0)\), and \((3,2)\). Lower panel: even
\((\ell+m)\) modes, \((\ell,m)=(1,1),(2,0),(2,2),(3,1)\), and
\((3,3)\). In the narrow band \(|b|\lesssim 5^\circ\) corresponding to
the adopted disk thickness, the even modes remain nearly constant at
\(O(1)\), whereas the odd modes are strongly suppressed.}
  \label{fig:legendre_b_profiles}
\end{figure}

The remaining latitude dependence is controlled by the parity under
reflection about the Galactic midplane, \(b\to -b\). The function
\(E^{\ell}_{m0}(b,0)\) is odd for odd \(\ell+m\) and even for even
\(\ell+m\). Thus the odd \((\ell+m)\) modes vanish at the midplane,
whereas the even \((\ell+m)\) modes remain \(O(1)\) and vary only
weakly within the thin disk (see Fig.~\ref{fig:legendre_b_profiles}).
It is convenient to define the function values on the midplane as
\begin{equation}
G^{\ell}_{m}\equiv E^{\ell}_{m0}(0,0)
=\sqrt{\frac{4\pi}{2\ell+1}}\,
Y_{\ell m}^\ast\!\left(\tfrac{\pi}{2},0\right),
\label{eq:G_def}
\end{equation}
which are explicitly given by
\begin{equation}
G^{\ell}_{m}=
\begin{cases}
0, & \ell+m \ \text{odd},\\[6pt]
(-1)^{(\ell+m)/2}
  \tfrac{(\ell+|m|-1)!!}{(\ell-|m|)!!}
  \sqrt{\tfrac{(\ell-|m|)!}{(\ell+|m|)!}}, & \ell+m \ \text{even}.
\end{cases}
\label{eq:G_closed}
\end{equation}
Numerical values for \(\ell\leq 3\) are listed in Table~\ref{tab:Gcoeffs}.
{In particular, the axisymmetric dipole has \(G^1_0=0\), so the
coefficient \(a_{10}\) is strongly suppressed in the midplane limit.
By contrast, \(G^2_0=-1/2\), making the quadrupolar coefficient
\(a_{20}\) the leading accessible axisymmetric mode in the present
setup.}

Combining the azimuthal factorization in Eq.~\eqref{eq:phi_fact} with
the weak \(b\)-dependence of \(E^{\ell}_{m0}(b,0)\) for the even-parity
modes in the disk-limited band \(|b|\lesssim5^\circ\), we use the
working approximation
\begin{equation}
E^{\ell}_{m0}(b,\varphi)\simeq G^{\ell}_{m}\,e^{-im\varphi}.
\label{eq:E_approx}
\end{equation}
For the low multipoles considered here, this approximation captures the
leading midplane contribution. Residual latitude-dependent corrections
are associated with the finite thickness of the disk and are treated as
subleading effects in the present analysis.

%%%%%%%%%%%%%%%%%%%%%%%%%%%%%%%%%%%%%%%%%%%
%%%%%%%%%%%%%%%%%%%%%%%%%%%%%%%%%%%%%%%%%%%

\renewcommand{\arraystretch}{1.4}
\begin{table}[t]
\centering
\caption{Nonvanishing coefficients \(G^\ell_m\) for low multipoles
(\(\ell \le 3\)). Exact expressions and approximate numerical values are
shown. Only \(m\ge 0\) are listed. For later reference, the last column
gives the \(1\sigma\) shot-noise error \(\Delta a_{\ell m}\), evaluated
from Eq.~\eqref{eq:alm_var} with the fiducial sample size
\(M_{\mathrm{tot}}=500\) and assuming uniform source counts across
longitude bins (\(F=1\)).}
\begin{tabular}{@{\hspace{0.8em}}c@{\hspace{1.5em}}c@{\hspace{1.5em}}c@{\hspace{1.5em}}c@{\hspace{0.8em}}}
\hline\hline
\(\ell\) & \(m\) & \(G^\ell_m\) (exact / approx) & \(\Delta a_{\ell m}\)  \\ \hline
1 & 1 & \(-1/\sqrt{2}\) \;\;/\;\; \(-0.707\) & \(0.0178\) \\
2 & 0 & \(-1/2\) \;\;/\;\; \(-0.500\) & \(0.0252\) \\
2 & 2 & \(3/(2\sqrt{6})\) \;\;/\;\; \(0.612\) & \(0.0206\) \\
3 & 1 & \(\sqrt{3}/4\) \;\;/\;\; \(0.433\) & \(0.0292\) \\
3 & 3 & \(-\sqrt{5}/4\) \;\;/\;\; \(-0.559\) & \(0.0226\) \\
\hline\hline
\end{tabular}
\label{tab:Gcoeffs}
\end{table}
%%%%%%%%%%%%%%%%%%%%%%%%%%%%%%%%%%%%%%%%%%%
%%%%%%%%%%%%%%%%%%%%%%%%%%%%%%%%%%%%%%%%%%%
\subsection{Extraction of coefficients}
\label{sec:extraction}

We now describe how to recover the harmonic coefficients \(a_{\ell m}\)
from the observed distribution of binaries in the disk-limited setup.
{The estimator introduced below is the disk-limited counterpart of
the patch-level observable in Eq.~\eqref{eq:patch_estimator}. For the
low multipoles considered here, Eq.~\eqref{eq:E_approx} reduces the
remaining sky dependence to a Fourier series in Galactic longitude, and
the estimator extracts its Fourier amplitudes.}

As discussed in Sec.~\ref{sec:kernels}, the projection kernels reduce to
\(E^{\ell}_{m0}(b,\varphi)\simeq G^\ell_m e^{-im\varphi}\), so that a
Fourier analysis along Galactic longitude \(\varphi\) provides direct
access to the coefficients \(a_{\ell m}\) with nonzero \(G^\ell_m\).
This can be regarded as the disk-limited version of the second step of
the inversion discussed in Sec.~\ref{sec:seg}, with the full-sky
projection replaced by a one-dimensional Fourier projection.

We divide Galactic longitude into \(N\) uniform bins of width
\(\Delta\varphi=2\pi/N\), centered at \(\varphi_k=2\pi k/N\)
(\(k=0,1,\dots,N-1\)). In each bin, following the construction in
Sec.~\ref{sec:patch}, we introduce the sky-limited estimator
\begin{equation}
S^\ell_{0}(\varphi_k)
= \frac{1}{M_k}\sum_{i=1}^{M_k} Y_{\ell 0}^\ast(I_{i}) ,
\label{eq:Sl_def}
\end{equation}
where \(M_k\) is the number of binaries in the \(k\)th bin and \(I_i\)
denotes the inclination of the \(i\)th binary in the bin. Note that, for
\(s=0\), we do not need the polarization angle \(\psi_i\).

From Eq.~\eqref{eq:E_approx}, the expectation value of
\(S^\ell_{0}(\varphi_k)\) can be expanded as
\begin{equation}
\mathbb{E} \lkk S^\ell_{0}(\varphi_k) \rkk
= \sum_{m} a_{\ell m} \,
G^\ell_m \, e^{-im\varphi_k} ,
\label{eq:Sl0}
\end{equation}
which is a Fourier series in \(\varphi_k\) with coefficients
proportional to the desired \(a_{\ell m}\). Modes with \(G^\ell_m=0\)
are not accessible in the midplane approximation.

As an illustration, for \(\ell=2\), only \(m=0\) and \(\pm 2\)
contribute:
\begin{equation}
\mathbb{E} \lkk S^2_{0}(\varphi_k)\rkk
= -\frac{1}{2}\,a_{20}
+ \frac{3}{2\sqrt{6}} \bigl(a_{22}\,e^{-2i\varphi_k}
+ a_{2,-2}\,e^{2i\varphi_k}\bigr).
\label{eq:S20_example}
\end{equation}
Analogous expressions for \(\ell=1\) and \(\ell=3\) can be derived in
the same way (see Table~\ref{tab:Gcoeffs} for the corresponding
\(G^\ell_m\) coefficients).

Since the longitude bins are uniformly spaced, and \(N\) is taken to be
sufficiently larger than the maximum \(|m|\) of interest, the relevant
discrete Fourier modes are orthogonal over the adopted grid. We can
therefore define a disk-limited estimator \(B_{\ell m}\) for the modes
with \(\ell+m\) even, equivalently \(G^\ell_m\ne0\), by
\begin{equation}
 B_{\ell m}
= \frac{1}{G^\ell_m\,N}\sum_{k=0}^{N-1}
   S^\ell_{0}(\varphi_k)\,e^{\,i m\varphi_k} ,
\label{eq:alm_estimator}
\end{equation}
which plays the same role as \(A_{\ell m}\) in Sec.~\ref{sec:seg}, but
with the full-sky projection replaced by a Fourier projection along
Galactic longitude.

It is straightforward to show that
\begin{equation}
\mathbb{E}[B_{\ell m}]=a_{\ell m}.
\end{equation}
Thus, for a narrow disk sample, we can construct unbiased estimators for
the coefficients with even parity (\(\ell+m\) even), within the
midplane approximation.

\section{Shot noise}\label{sec:shot}

{In the previous section we identified which harmonic coefficients can be
extracted in the disk-limited geometry. We now estimate how accurately
these coefficients can be measured with a finite number of resolved
DWDs.}
The dominant statistical uncertainty is shot noise, arising from the
finite number of binaries in each longitude bin. We evaluate this
uncertainty for the disk-limited estimators \(B_{\ell m}\) introduced in
Sec.~\ref{sec:extraction}. Throughout we work in the weak-signal regime,
defined by \(|a_{\ell m}|\ll 1\), where the local orientation
distribution \(Q_{\rm loc}\) is close to isotropic (see
Ref.~\cite{Jammalamadaka2019} for a more general treatment). The
resulting estimates provide a benchmark for the detection limits of
orientation patterns.

\subsection{Shot-noise variance of the estimator}\label{sec:var}

The starting point is the variance of the longitude-bin averages
\(S^\ell_{0}(\varphi_k)\) defined in Eq.~(\ref{eq:Sl_def}). In the
weak-signal approximation, one finds
\begin{equation}
\mathrm{Var}\!\left[S^\ell_{0}(\varphi_k)\right]
= \frac{1}{M_k}\left\langle |Y_{\ell 0}(I)|^2 \right\rangle_{\rm iso}
= \frac{1}{4\pi\,M_k},
\label{eq:S_var}
\end{equation}
where \(M_k\) is the number of binaries in the \(k\)th longitude bin.
The final expression follows from the isotropic average
\(\langle |Y_{\ell 0}|^2\rangle_{\rm iso}=1/4\pi\) in the weak-signal
limit.

From Eqs.~(\ref{eq:alm_estimator}) and (\ref{eq:S_var}), and assuming
that shot-noise fluctuations in different longitude bins are independent,
we obtain
\begin{equation}
\mathrm{Var}\!\left[B_{\ell m}\right]
= \frac{1}{|G^{\ell}_{m}|^2 N^2}
\sum_{k=0}^{N-1} \mathrm{Var}\!\left[S^\ell_{0}(\varphi_k)\right].
\label{eq:alm_var_step}
\end{equation}
Carrying out the sum leads to
\begin{equation}
\mathrm{Var}\!\left[B_{\ell m}\right]
= \frac{1}{4\pi\,|G^{\ell}_{m}|^2\,M_{\mathrm{tot}}}\,F^2 ,
\label{eq:alm_var}
\end{equation}
where
\begin{equation}
M_{\mathrm{tot}} \equiv \sum_{k=0}^{N-1} M_k
\label{eq:M_total}
\end{equation}
is the total number of binaries, and
\begin{equation}
F \equiv \left[\frac{1}{N^2}\sum_{k=0}^{N-1}
\frac{M_{\mathrm{tot}}}{M_k}\right]^{1/2}
\;\;\ge 1
\label{eq:F_factor}
\end{equation}
is the degradation factor due to the longitudinal non-uniformity of
source counts. The factor \(F\) equals unity for uniform bin counts and
exceeds unity whenever \(M_k\ne M_{\mathrm{tot}}/N\).

Equation~(\ref{eq:alm_var}) shows that the error scales as
\(M_{\mathrm{tot}}^{-1/2}\), as expected for Poisson noise. The variance
is weighted by the geometric factor \(|G^{\ell}_{m}|^{-2}\), and
detectability is therefore mode dependent. As a reference,
Table~\ref{tab:Gcoeffs} also lists the corresponding shot-noise errors
\(\Delta a_{\ell m}\) computed from Eq.~(\ref{eq:alm_var}) for the
fiducial sample size \(M_{\mathrm{tot}}=500\) and \(F=1\).

In discussing the covariance structure of the shot-noise errors, it is
useful to distinguish the roles of \(\ell\) and \(m\). For different
values of \(\ell\), the orthogonality of the inclination-dependent
functions \(Y_{\ell0}(I)\) ensures that the corresponding modes do not
correlate in the weak-signal limit. For the same \(\ell\) but different
\(m\), the cancellation follows from the orthogonality of the discrete
Fourier sum in longitude, provided that the bin counts \(M_k\) are
uniform. If the counts are non-uniform, however, this orthogonality is
no longer exact, and weak off-diagonal covariances may arise between
different \(m\) modes.

\subsection{Illustration with longitudinal count contrast}

\begin{figure}[t]
\centering
\includegraphics[width=0.95\linewidth]{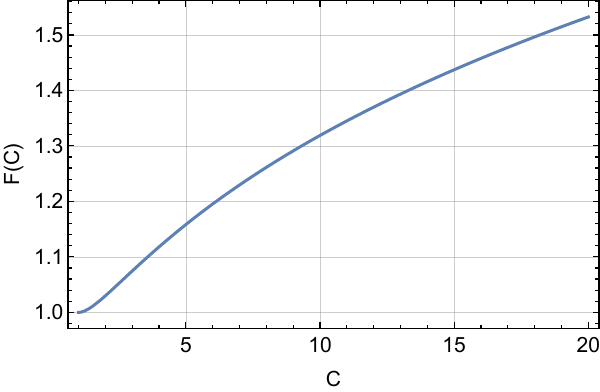}
\caption{Degradation factor \(F\) as a function of the longitudinal
source-count contrast \(C=(1+\varepsilon)/(1-\varepsilon)\). The
dependence is weak: even \(C=19\) (\(\varepsilon=0.9\)) gives only
\(F\simeq 1.5\).}
\label{fig:FvsC}
\end{figure}

As an illustrative model for the non-uniformity of source counts along
Galactic longitude, suppose the expected counts follow
\begin{equation}
M_k\propto 1+\varepsilon\cos\varphi_k ,
\label{jkm}
\end{equation}
with \(0\le \varepsilon<1\). The ratio of maximum to minimum counts is
then
\begin{equation}
C \equiv \frac{1+\varepsilon}{1-\varepsilon},
\end{equation}
which provides a convenient measure of the longitudinal count contrast
along the Galactic plane. Replacing the sum in Eq.~(\ref{eq:F_factor})
with an integral, we obtain
\begin{equation}
F = (1-\varepsilon^2)^{-1/4}
  = \frac{\sqrt{1+C}}{\sqrt{2}\,C^{1/4}} .
\label{eq:F_example}
\end{equation}
This form shows that \(F\) grows only slowly with increasing contrast.
Figure~\ref{fig:FvsC} presents \(F\) as a function of \(C\). For example,
\(C=3\) (\(\varepsilon=0.5\)) gives \(F\simeq 1.08\), while \(C=9\)
(\(\varepsilon=0.8\)) gives \(F\simeq 1.29\). Thus even large
source-count contrasts produce only moderate degradation.

The same non-uniformity of source counts can also induce off-diagonal
covariances between different \(m\)-modes.
{This covariance arises because non-uniform source counts make the
shot-noise variance longitude dependent. If all bins have the same
count \(M_k\), the Fourier modes remain orthogonal in the noise
covariance. Once \(M_k\) varies with \(\varphi_k\), the Fourier
projection samples this variance pattern, and different \(m\)-modes can
become weakly correlated.}
In the present low-multipole disk case, the coefficients \(G^\ell_m\)
are real. Applying the integral approximation to the analytical model
in Eq.~\eqref{jkm}, the off-diagonal covariance for \(m\ne m'\) is
\begin{equation}
\mathrm{Cov}\!\left[B_{\ell m},B_{\ell m'}\right]
=\frac{1}{4\pi\,G^\ell_m G^\ell_{m'}\,M_{\rm tot}}\,
\frac{r(\varepsilon)^{\,|m-m'|}}{\sqrt{1-\varepsilon^2}} ,
\end{equation}
where
\begin{equation}
r(\varepsilon)=\frac{1-\sqrt{\,1-\varepsilon^2\,}}{\varepsilon}
=\frac{1}{2}\varepsilon+O(\varepsilon^3) .
\end{equation}
Numerically, \(r\simeq0.268\) for \(\varepsilon=0.5\) and
\(r\simeq0.333\) for \(\varepsilon=0.8\). The corresponding correlation
coefficient is
\begin{equation}
\rho_{mm'} \equiv 
\frac{\mathrm{Cov}[B_{\ell m},B_{\ell m'}]}
{\sqrt{\mathrm{Var}[B_{\ell m}]\,\mathrm{Var}[B_{\ell m'}]}}
= \mathrm{sgn}\!\big(G^\ell_m G^\ell_{m'}\big)\,
r(\varepsilon)^{\,|m-m'|} ,
\end{equation}
so that the magnitude of the correlation decays geometrically as
\[
|\rho_{mm'}|=r(\varepsilon)^{\,|m-m'|}.
\]

%%%%%%%%%%%%%%%%%%%%%%%%%%%%%%%%%%%%%
%%%%%%%%%%%%%%%%%%%%%%%%%%%%%%%%

\subsection{Parameter-estimation errors}

In addition to shot noise, parameter-estimation errors in the
source-frame orientation angles can introduce extra fluctuations in the
recovered coefficients. For the low-multipole disk-limited estimators
used above, only the \(s=0\) component enters, and the dominant effect is
the error in the inclination \(I\). More generally, the polarization
angle \(\psi\) also enters once \(s\ne0\) components are used. For each
binary, such angular errors are roughly of order \(\rho^{-1}\), where
\(\rho\) is its signal-to-noise ratio. Therefore, the induced variance
of the recovered coefficients can be estimated as
\begin{equation}
\mathrm{Var}_{\rm param}\!\left[B_{\ell m}\right]
\sim \frac{1}{M_{\mathrm{tot}}}\,\frac{\kappa_{\ell m}}{\rho_t^2},
\label{eq:param_noise}
\end{equation}
where \(\rho_t\) is the typical single-source SNR and
\(\kappa_{\ell m}\) is a geometry factor of order unity
\cite{Cutler:1997ta,Takahashi:2002ky}. In most practical cases
(\(\rho_t\gg 1\)), the contribution of Eq.~(\ref{eq:param_noise}) is
subdominant compared to the shot-noise variance of Eq.~(\ref{eq:alm_var}).

\section{Discussion}
We now discuss several limitations of the present local, low-multipole
analysis and possible extensions.

\subsection{Limitations of the local approximation}\label{sec:disk_limits}

In our analysis we have restricted the sample to binaries within
\(d_{\max}=4\,\mathrm{kpc}\) and focused on the effective local
orientation distribution \(Q_{\rm loc}(\hat{\bm l})\) averaged over the
selected sample. Thus we do not attempt to resolve the spatial
dependence of the orientation distribution within the selected volume. This treatment is motivated by
the localized spatial extent of the survey region in our Galaxy.

The extracted coefficients \(a_{\ell m}\) should therefore be
interpreted as sample-averaged local moments over the restricted survey
volume. If sufficiently strong patterns are detected in the local
sample, the method could be extended to investigate the spatial
dependence of the orientation distribution itself, using a larger number
of DWDs over a wider survey volume. Such an analysis would open a new
line of inquiry, potentially offering insights into Galactic star
formation and binary evolution processes.

\subsection{Higher multipoles ($\ell \geq 4$)}\label{sec:higher_l}

Our method can in principle be extended to arbitrary multipoles.
In the low-multipole analysis with \(\ell\le3\), the fourfold
polarization degeneracy leaves only the \(s=0\) term. For
\(\ell \geq 4\), additional source-frame harmonics
(\(s=\pm4,\pm8,\dots\), with \(|s|\le\ell\)) contribute alongside the
\(s=0\) term. Because the \(s\) and \(-s\) components are related by
complex conjugation for a real distribution, the independent degrees of
freedom can be restricted to \(s\ge0\), whose number grows roughly in
proportion to \(\ell\).

In the weak-signal limit, the shot-noise contributions of different
\(s\)-modes are uncorrelated. With the notation of Sec.~\ref{sec:var}
and \(\langle\cdot\rangle_{\rm iso}\) denoting isotropic averaging, we
have, for example,
\[
\big\langle\, Y_{4 0}(I)\; Y_{4 4}(I,\psi)^{*} \big\rangle_{\rm iso}=0.
\]

\subsection{Odd $(\ell+m)$ modes}\label{sec:odd_modes}

As is evident from our expansion, modes with odd \((\ell+m)\) vanish
exactly on the Galactic midplane (\(b=0\)). This is a direct consequence
of the parity symmetry of the spherical harmonics. In particular, the
axisymmetric dipole \(a_{10}\) belongs to this class, because
\(G^1_0=0\). To detect such modes, one must rely on asymmetries above
and below the Galactic plane, or equivalently on the finite vertical
thickness of the disk. A similar situation was found in the
bulge-oriented analysis of Ref.~\cite{Seto:2024wrw}.

Because the odd contributions arise only through the finite disk
thickness, they are particularly susceptible to shot noise. In practice,
the effective signal-to-noise ratio is suppressed by a factor
proportional to the characteristic angular width \(b_t\sim 0.1\)
compared to the even modes (see Ref.~\cite{Seto:2024wrw} for a related
argument). With a total sample size of \(M_{\mathrm{tot}}\sim 500\), it
would be statistically difficult to detect the odd modes in the present
setup.

\subsection{Other technical considerations}\label{sec:others}

Our extraction scheme employs \(N\) longitude bins, so that the recovery
of \(a_{\ell m}\) is effectively a discrete Fourier transform rather
than a continuous one. This introduces two possible corrections: (i) a
mild sinc-type suppression because each bin averages over a finite
longitude interval, and (ii) potential aliasing if \(N\) is too small.
In practice these effects are negligible for the present analysis. Since
we are only interested in multipoles with \(\ell \le 3\), and hence
\(|m|\le 3\), choosing \(N \gtrsim 20\) makes aliasing irrelevant, while
the bin-averaging correction remains well below the statistical errors
from shot noise.

\section{Summary}

In this work, we presented a method to quantify the orientation
distribution of nearby (\(\lesssim 4\) kpc) DWDs using a
spherical-harmonic expansion. The selected sample corresponds to about
5\% of the resolved Galactic DWD population
(\(M_{\mathrm{tot}}\sim 500\)). We described the target quantity as the effective local distribution
\(Q_{\rm loc}(\hat{\bm l})\) of orbital angular momentum directions in
Galactic coordinates, averaged over the selected sample. The extracted
coefficients \(a_{\ell m}\) should therefore be interpreted as
sample-averaged local moments of the orientation distribution within the
selected survey volume.

Using the Wigner-\(D\) representation, we related this Galactic-frame
distribution to the source-frame angles \(I\) and \(\psi\), and
explicitly accounted for the fourfold degeneracy in the polarization
angle. For multipoles up to \(\ell \leq 3\), this degeneracy leaves only
the \(s=0\) components. In the disk-limited regime, the kernel values
\(G^\ell_m\) at the Galactic midplane determine the sensitivity: even
\((\ell+m)\) modes remain finite, while odd modes vanish by symmetry.
In particular, the axisymmetric dipole \(a_{10}\) is suppressed in the
midplane limit, whereas the quadrupolar coefficient \(a_{20}\) remains
accessible and provides the leading axisymmetric probe of whether the orbital angular momentum vectors preferentially point
toward the Galactic poles or lie in the Galactic plane.

We used an ideal full-sky case only as a reference for the inverse
problem, and then adapted the extraction to the belt-like sky
distribution of nearby disk DWDs. In this disk-limited geometry, the
remaining longitude dependence of the projection kernels allows the
accessible coefficients \(a_{\ell m}\) to be extracted by a Fourier
transform along Galactic longitude. This step contrasts with earlier
bulge-focused studies, where the sky distribution is confined to a
localized region \cite{Seto:2024odc}.

{After identifying the accessible modes in the disk-limited geometry,
we evaluated the statistical precision with which they can be measured
from a finite sample of resolved DWDs.} For
\(M_{\mathrm{tot}}\sim 500\), the typical uncertainty for accessible
even modes with \(\ell\le 3\) is of order
\(\Delta a_{\ell m}\sim0.02\). The error scales approximately as
\(M_{\mathrm{tot}}^{-1/2}\), with only a moderate degradation from the
longitudinal non-uniformity of source counts, represented by the factor
\(F\). In contrast, odd \((\ell+m)\) modes are unlikely to be detected
in the present setup, since their weak residual signals from the finite
disk thickness are easily buried in shot noise.

Future applications to larger survey regions may allow us to probe the
spatial dependence of the orientation distribution in the Galactic disk.

\section*{Acknowledgments}
The authors thank H. Omiya for valuable discussions.
This work was supported in part by JSPS KAKENHI Grant-in-Aid for
Scientific Research Nos.~23K03385 and 26H00834 (NS), and by the
Ministry of Science and Technology of Taiwan, Republic of China, under
Grant No.~113-2112-M-032-011 (GCL).
NS also acknowledges the hospitality of the Munich Institute for
Astro- and Particle Physics (MIAPbP) in Garching, where part of this
work was carried out.

\bibliographystyle{apsrev4-2}
\bibliography{ref2} 

\begin{thebibliography}{22}%
\makeatletter
\providecommand \@ifxundefined [1]{%
 \@ifx{#1\undefined}
}%
\providecommand \@ifnum [1]{%
 \ifnum #1\expandafter \@firstoftwo
 \else \expandafter \@secondoftwo
 \fi
}%
\providecommand \@ifx [1]{%
 \ifx #1\expandafter \@firstoftwo
 \else \expandafter \@secondoftwo
 \fi
}%
\providecommand \natexlab [1]{#1}%
\providecommand \enquote  [1]{``#1''}%
\providecommand \bibnamefont  [1]{#1}%
\providecommand \bibfnamefont [1]{#1}%
\providecommand \citenamefont [1]{#1}%
\providecommand \href@noop [0]{\@secondoftwo}%
\providecommand \href [0]{\begingroup \@sanitize@url \@href}%
\providecommand \@href[1]{\@@startlink{#1}\@@href}%
\providecommand \@@href[1]{\endgroup#1\@@endlink}%
\providecommand \@sanitize@url [0]{\catcode `\\12\catcode `\$12\catcode
  `\&12\catcode `\#12\catcode `\^12\catcode `\_12\catcode `\%12\relax}%
\providecommand \@@startlink[1]{}%
\providecommand \@@endlink[0]{}%
\providecommand \url  [0]{\begingroup\@sanitize@url \@url }%
\providecommand \@url [1]{\endgroup\@href {#1}{\urlprefix }}%
\providecommand \urlprefix  [0]{URL }%
\providecommand \Eprint [0]{\href }%
\providecommand \doibase [0]{https://doi.org/}%
\providecommand \selectlanguage [0]{\@gobble}%
\providecommand \bibinfo  [0]{\@secondoftwo}%
\providecommand \bibfield  [0]{\@secondoftwo}%
\providecommand \translation [1]{[#1]}%
\providecommand \BibitemOpen [0]{}%
\providecommand \bibitemStop [0]{}%
\providecommand \bibitemNoStop [0]{.\EOS\space}%
\providecommand \EOS [0]{\spacefactor3000\relax}%
\providecommand \BibitemShut  [1]{\csname bibitem#1\endcsname}%
\let\auto@bib@innerbib\@empty
%</preamble>
\bibitem [{\citenamefont {{Chang}}(1929)}]{1929AJ.....40...11C}%
  \BibitemOpen
  \bibfield  {author} {\bibinfo {author} {\bibfnamefont {Y.~C.}\ \bibnamefont
  {{Chang}}},\ }\href {https://doi.org/10.1086/104946} {\bibfield  {journal}
  {\bibinfo  {journal} {AJ}\ }\textbf {\bibinfo {volume} {40}},\ \bibinfo
  {pages} {11} (\bibinfo {year} {1929})}\BibitemShut {NoStop}%
\bibitem [{\citenamefont {{Agati}}\ \emph {et~al.}(2015)\citenamefont
  {{Agati}}, \citenamefont {{Bonneau}}, \citenamefont {{Jorissen}},
  \citenamefont {{Souli{\'e}}}, \citenamefont {{Udry}}, \citenamefont
  {{Verhas}},\ and\ \citenamefont {{Dommanget}}}]{2015A&A...574A...6A}%
  \BibitemOpen
  \bibfield  {author} {\bibinfo {author} {\bibfnamefont {J.~L.}\ \bibnamefont
  {{Agati}}}, \bibinfo {author} {\bibfnamefont {D.}~\bibnamefont {{Bonneau}}},
  \bibinfo {author} {\bibfnamefont {A.}~\bibnamefont {{Jorissen}}}, \bibinfo
  {author} {\bibfnamefont {E.}~\bibnamefont {{Souli{\'e}}}}, \bibinfo {author}
  {\bibfnamefont {S.}~\bibnamefont {{Udry}}}, \bibinfo {author} {\bibfnamefont
  {P.}~\bibnamefont {{Verhas}}},\ and\ \bibinfo {author} {\bibfnamefont
  {J.}~\bibnamefont {{Dommanget}}},\ }\href
  {https://doi.org/10.1051/0004-6361/201323056} {\bibfield  {journal} {\bibinfo
   {journal} {A\&A}\ }\textbf {\bibinfo {volume} {574}},\ \bibinfo {eid} {A6}
  (\bibinfo {year} {2015})},\ \Eprint {https://arxiv.org/abs/1411.4919}
  {arXiv:1411.4919 [astro-ph.SR]} \BibitemShut {NoStop}%
\bibitem [{\citenamefont {Arenou}\ \emph {et~al.}(2023)\citenamefont {Arenou},
  \citenamefont {Babusiaux}, \citenamefont {Barstow}, \citenamefont {Faigler},
  \citenamefont {Jorissen}, \citenamefont {Kervella}, \citenamefont {Mazeh},
  \citenamefont {Mowlavi}, \citenamefont {Panuzzo}, \citenamefont {Sahlmann}
  \emph {et~al.}}]{arenou2023gaia}%
  \BibitemOpen
  \bibfield  {author} {\bibinfo {author} {\bibfnamefont {F.}~\bibnamefont
  {Arenou}}, \bibinfo {author} {\bibfnamefont {C.}~\bibnamefont {Babusiaux}},
  \bibinfo {author} {\bibfnamefont {M.~A.}\ \bibnamefont {Barstow}}, \bibinfo
  {author} {\bibfnamefont {S.}~\bibnamefont {Faigler}}, \bibinfo {author}
  {\bibfnamefont {A.}~\bibnamefont {Jorissen}}, \bibinfo {author}
  {\bibfnamefont {P.}~\bibnamefont {Kervella}}, \bibinfo {author}
  {\bibfnamefont {T.}~\bibnamefont {Mazeh}}, \bibinfo {author} {\bibfnamefont
  {N.}~\bibnamefont {Mowlavi}}, \bibinfo {author} {\bibfnamefont
  {P.}~\bibnamefont {Panuzzo}}, \bibinfo {author} {\bibfnamefont
  {J.}~\bibnamefont {Sahlmann}}, \emph {et~al.},\ }\href@noop {} {\bibfield
  {journal} {\bibinfo  {journal} {Astronomy \& Astrophysics}\ }\textbf
  {\bibinfo {volume} {674}},\ \bibinfo {pages} {A34} (\bibinfo {year}
  {2023})}\BibitemShut {NoStop}%
\bibitem [{\citenamefont {{Tan}}\ \emph {et~al.}(2023)\citenamefont {{Tan}},
  \citenamefont {{Parker}}, \citenamefont {{Zijlstra}}, \citenamefont
  {{Ritter}},\ and\ \citenamefont {{Rees}}}]{2023ApJ...951L..44T}%
  \BibitemOpen
  \bibfield  {author} {\bibinfo {author} {\bibfnamefont {S.}~\bibnamefont
  {{Tan}}}, \bibinfo {author} {\bibfnamefont {Q.~A.}\ \bibnamefont {{Parker}}},
  \bibinfo {author} {\bibfnamefont {A.~A.}\ \bibnamefont {{Zijlstra}}},
  \bibinfo {author} {\bibfnamefont {A.}~\bibnamefont {{Ritter}}},\ and\
  \bibinfo {author} {\bibfnamefont {B.}~\bibnamefont {{Rees}}},\ }\href
  {https://doi.org/10.3847/2041-8213/acdbcd} {\bibfield  {journal} {\bibinfo
  {journal} {ApJ}\ }\textbf {\bibinfo {volume} {951}},\ \bibinfo {eid} {L44}
  (\bibinfo {year} {2023})},\ \Eprint {https://arxiv.org/abs/2307.07140}
  {arXiv:2307.07140 [astro-ph.GA]} \BibitemShut {NoStop}%
\bibitem [{\citenamefont {{Poisson}}\ and\ \citenamefont
  {{Will}}(2014)}]{2014grav.book.....P}%
  \BibitemOpen
  \bibfield  {author} {\bibinfo {author} {\bibfnamefont {E.}~\bibnamefont
  {{Poisson}}}\ and\ \bibinfo {author} {\bibfnamefont {C.~M.}\ \bibnamefont
  {{Will}}},\ }\href@noop {} {\emph {\bibinfo {title} {{Gravity}}}}\ (\bibinfo
  {year} {2014})\BibitemShut {NoStop}%
\bibitem [{\citenamefont {Cornish}\ and\ \citenamefont
  {Larson}(2003)}]{Cornish:2003vj}%
  \BibitemOpen
  \bibfield  {author} {\bibinfo {author} {\bibfnamefont {N.~J.}\ \bibnamefont
  {Cornish}}\ and\ \bibinfo {author} {\bibfnamefont {S.~L.}\ \bibnamefont
  {Larson}},\ }\href {https://doi.org/10.1103/PhysRevD.67.103001} {\bibfield
  {journal} {\bibinfo  {journal} {Phys. Rev. D}\ }\textbf {\bibinfo {volume}
  {67}},\ \bibinfo {pages} {103001} (\bibinfo {year} {2003})},\ \Eprint
  {https://arxiv.org/abs/astro-ph/0301548} {arXiv:astro-ph/0301548}
  \BibitemShut {NoStop}%
\bibitem [{\citenamefont {Seoane}\ \emph {et~al.}(2023)\citenamefont {Seoane}
  \emph {et~al.}}]{LISA:2022yao}%
  \BibitemOpen
  \bibfield  {author} {\bibinfo {author} {\bibfnamefont {P.~A.}\ \bibnamefont
  {Seoane}} \emph {et~al.} (\bibinfo {collaboration} {LISA}),\ }\href
  {https://doi.org/10.1007/s41114-022-00041-y} {\bibfield  {journal} {\bibinfo
  {journal} {Living Rev. Rel.}\ }\textbf {\bibinfo {volume} {26}},\ \bibinfo
  {pages} {2} (\bibinfo {year} {2023})},\ \Eprint
  {https://arxiv.org/abs/2203.06016} {arXiv:2203.06016 [gr-qc]} \BibitemShut
  {NoStop}%
\bibitem [{\citenamefont {Nissanke}\ \emph {et~al.}(2012)\citenamefont
  {Nissanke}, \citenamefont {Vallisneri}, \citenamefont {Nelemans},\ and\
  \citenamefont {Prince}}]{Nissanke:2012eh}%
  \BibitemOpen
  \bibfield  {author} {\bibinfo {author} {\bibfnamefont {S.}~\bibnamefont
  {Nissanke}}, \bibinfo {author} {\bibfnamefont {M.}~\bibnamefont
  {Vallisneri}}, \bibinfo {author} {\bibfnamefont {G.}~\bibnamefont
  {Nelemans}},\ and\ \bibinfo {author} {\bibfnamefont {T.~A.}\ \bibnamefont
  {Prince}},\ }\href {https://doi.org/10.1088/0004-637X/758/2/131} {\bibfield
  {journal} {\bibinfo  {journal} {Astrophys. J.}\ }\textbf {\bibinfo {volume}
  {758}},\ \bibinfo {pages} {131} (\bibinfo {year} {2012})},\ \Eprint
  {https://arxiv.org/abs/1201.4613} {arXiv:1201.4613 [astro-ph.GA]}
  \BibitemShut {NoStop}%
\bibitem [{\citenamefont {Korol}\ \emph {et~al.}(2017)\citenamefont {Korol},
  \citenamefont {Rossi}, \citenamefont {Groot}, \citenamefont {Nelemans},
  \citenamefont {Toonen},\ and\ \citenamefont {Brown}}]{Korol:2017qcx}%
  \BibitemOpen
  \bibfield  {author} {\bibinfo {author} {\bibfnamefont {V.}~\bibnamefont
  {Korol}}, \bibinfo {author} {\bibfnamefont {E.~M.}\ \bibnamefont {Rossi}},
  \bibinfo {author} {\bibfnamefont {P.~J.}\ \bibnamefont {Groot}}, \bibinfo
  {author} {\bibfnamefont {G.}~\bibnamefont {Nelemans}}, \bibinfo {author}
  {\bibfnamefont {S.}~\bibnamefont {Toonen}},\ and\ \bibinfo {author}
  {\bibfnamefont {A.~G.~A.}\ \bibnamefont {Brown}},\ }\href
  {https://doi.org/10.1093/mnras/stx1285} {\bibfield  {journal} {\bibinfo
  {journal} {Mon. Not. Roy. Astron. Soc.}\ }\textbf {\bibinfo {volume} {470}},\
  \bibinfo {pages} {1894} (\bibinfo {year} {2017})},\ \Eprint
  {https://arxiv.org/abs/1703.02555} {arXiv:1703.02555 [astro-ph.HE]}
  \BibitemShut {NoStop}%
\bibitem [{\citenamefont {Lamberts}\ \emph {et~al.}(2019)\citenamefont
  {Lamberts}, \citenamefont {Blunt}, \citenamefont {Littenberg}, \citenamefont
  {Garrison-Kimmel}, \citenamefont {Kupfer},\ and\ \citenamefont
  {Sanderson}}]{Lamberts:2019nyk}%
  \BibitemOpen
  \bibfield  {author} {\bibinfo {author} {\bibfnamefont {A.}~\bibnamefont
  {Lamberts}}, \bibinfo {author} {\bibfnamefont {S.}~\bibnamefont {Blunt}},
  \bibinfo {author} {\bibfnamefont {T.~B.}\ \bibnamefont {Littenberg}},
  \bibinfo {author} {\bibfnamefont {S.}~\bibnamefont {Garrison-Kimmel}},
  \bibinfo {author} {\bibfnamefont {T.}~\bibnamefont {Kupfer}},\ and\ \bibinfo
  {author} {\bibfnamefont {R.~E.}\ \bibnamefont {Sanderson}},\ }\href
  {https://doi.org/10.1093/mnras/stz2834} {\bibfield  {journal} {\bibinfo
  {journal} {Mon. Not. Roy. Astron. Soc.}\ }\textbf {\bibinfo {volume} {490}},\
  \bibinfo {pages} {5888} (\bibinfo {year} {2019})},\ \Eprint
  {https://arxiv.org/abs/1907.00014} {arXiv:1907.00014 [astro-ph.HE]}
  \BibitemShut {NoStop}%
\bibitem [{\citenamefont {Hu}\ and\ \citenamefont {Wu}(2017)}]{Hu:2017mde}%
  \BibitemOpen
  \bibfield  {author} {\bibinfo {author} {\bibfnamefont {W.-R.}\ \bibnamefont
  {Hu}}\ and\ \bibinfo {author} {\bibfnamefont {Y.-L.}\ \bibnamefont {Wu}},\
  }\href {https://doi.org/10.1093/nsr/nwx116} {\bibfield  {journal} {\bibinfo
  {journal} {Natl. Sci. Rev.}\ }\textbf {\bibinfo {volume} {4}},\ \bibinfo
  {pages} {685} (\bibinfo {year} {2017})}\BibitemShut {NoStop}%
\bibitem [{\citenamefont {Luo}\ \emph {et~al.}(2016)\citenamefont {Luo} \emph
  {et~al.}}]{TianQin:2015yph}%
  \BibitemOpen
  \bibfield  {author} {\bibinfo {author} {\bibfnamefont {J.}~\bibnamefont
  {Luo}} \emph {et~al.} (\bibinfo {collaboration} {TianQin}),\ }\href
  {https://doi.org/10.1088/0264-9381/33/3/035010} {\bibfield  {journal}
  {\bibinfo  {journal} {Class. Quant. Grav.}\ }\textbf {\bibinfo {volume}
  {33}},\ \bibinfo {pages} {035010} (\bibinfo {year} {2016})},\ \Eprint
  {https://arxiv.org/abs/1512.02076} {arXiv:1512.02076 [astro-ph.IM]}
  \BibitemShut {NoStop}%
\bibitem [{\citenamefont {Huang}\ \emph {et~al.}(2020)\citenamefont {Huang},
  \citenamefont {Hu}, \citenamefont {Korol}, \citenamefont {Li}, \citenamefont
  {Liang}, \citenamefont {Lu}, \citenamefont {Wang}, \citenamefont {Yu},\ and\
  \citenamefont {Mei}}]{Huang:2020rjf}%
  \BibitemOpen
  \bibfield  {author} {\bibinfo {author} {\bibfnamefont {S.-J.}\ \bibnamefont
  {Huang}}, \bibinfo {author} {\bibfnamefont {Y.-M.}\ \bibnamefont {Hu}},
  \bibinfo {author} {\bibfnamefont {V.}~\bibnamefont {Korol}}, \bibinfo
  {author} {\bibfnamefont {P.-C.}\ \bibnamefont {Li}}, \bibinfo {author}
  {\bibfnamefont {Z.-C.}\ \bibnamefont {Liang}}, \bibinfo {author}
  {\bibfnamefont {Y.}~\bibnamefont {Lu}}, \bibinfo {author} {\bibfnamefont
  {H.-T.}\ \bibnamefont {Wang}}, \bibinfo {author} {\bibfnamefont
  {S.}~\bibnamefont {Yu}},\ and\ \bibinfo {author} {\bibfnamefont
  {J.}~\bibnamefont {Mei}},\ }\href
  {https://doi.org/10.1103/PhysRevD.102.063021} {\bibfield  {journal} {\bibinfo
   {journal} {Phys. Rev. D}\ }\textbf {\bibinfo {volume} {102}},\ \bibinfo
  {pages} {063021} (\bibinfo {year} {2020})},\ \Eprint
  {https://arxiv.org/abs/2005.07889} {arXiv:2005.07889 [astro-ph.HE]}
  \BibitemShut {NoStop}%
\bibitem [{\citenamefont {Takahashi}\ and\ \citenamefont
  {Seto}(2002)}]{Takahashi:2002ky}%
  \BibitemOpen
  \bibfield  {author} {\bibinfo {author} {\bibfnamefont {R.}~\bibnamefont
  {Takahashi}}\ and\ \bibinfo {author} {\bibfnamefont {N.}~\bibnamefont
  {Seto}},\ }\href {https://doi.org/10.1086/341483} {\bibfield  {journal}
  {\bibinfo  {journal} {Astrophys. J.}\ }\textbf {\bibinfo {volume} {575}},\
  \bibinfo {pages} {1030} (\bibinfo {year} {2002})},\ \Eprint
  {https://arxiv.org/abs/astro-ph/0204487} {arXiv:astro-ph/0204487}
  \BibitemShut {NoStop}%
\bibitem [{\citenamefont {Cutler}(1998)}]{Cutler:1997ta}%
  \BibitemOpen
  \bibfield  {author} {\bibinfo {author} {\bibfnamefont {C.}~\bibnamefont
  {Cutler}},\ }\href {https://doi.org/10.1103/PhysRevD.57.7089} {\bibfield
  {journal} {\bibinfo  {journal} {Phys. Rev. D}\ }\textbf {\bibinfo {volume}
  {57}},\ \bibinfo {pages} {7089} (\bibinfo {year} {1998})},\ \Eprint
  {https://arxiv.org/abs/gr-qc/9703068} {arXiv:gr-qc/9703068} \BibitemShut
  {NoStop}%
\bibitem [{\citenamefont {Seto}(2024{\natexlab{a}})}]{Seto:2024odc}%
  \BibitemOpen
  \bibfield  {author} {\bibinfo {author} {\bibfnamefont {N.}~\bibnamefont
  {Seto}},\ }\href {https://doi.org/10.1103/PhysRevD.109.103016} {\bibfield
  {journal} {\bibinfo  {journal} {Phys. Rev. D}\ }\textbf {\bibinfo {volume}
  {109}},\ \bibinfo {pages} {103016} (\bibinfo {year} {2024}{\natexlab{a}})},\
  \Eprint {https://arxiv.org/abs/2404.13313} {arXiv:2404.13313 [astro-ph.HE]}
  \BibitemShut {NoStop}%
\bibitem [{\citenamefont {Seto}(2024{\natexlab{b}})}]{Seto:2024wrw}%
  \BibitemOpen
  \bibfield  {author} {\bibinfo {author} {\bibfnamefont {N.}~\bibnamefont
  {Seto}},\ }\href {https://doi.org/10.1103/PhysRevD.110.123003} {\bibfield
  {journal} {\bibinfo  {journal} {Phys. Rev. D}\ }\textbf {\bibinfo {volume}
  {110}},\ \bibinfo {pages} {123003} (\bibinfo {year} {2024}{\natexlab{b}})},\
  \Eprint {https://arxiv.org/abs/2411.12961} {arXiv:2411.12961 [astro-ph.GA]}
  \BibitemShut {NoStop}%
\bibitem [{\citenamefont {Edmonds}(1957)}]{Edmonds1957}%
  \BibitemOpen
  \bibfield  {author} {\bibinfo {author} {\bibfnamefont {A.~R.}\ \bibnamefont
  {Edmonds}},\ }\href@noop {} {\emph {\bibinfo {title} {Angular Momentum in
  Quantum Mechanics}}}\ (\bibinfo  {publisher} {Princeton University Press},\
  \bibinfo {year} {1957})\BibitemShut {NoStop}%
\bibitem [{\citenamefont {Thorne}(1980)}]{Thorne:1980rev}%
  \BibitemOpen
  \bibfield  {author} {\bibinfo {author} {\bibfnamefont {K.~S.}\ \bibnamefont
  {Thorne}},\ }\href {https://doi.org/10.1103/RevModPhys.52.299} {\bibfield
  {journal} {\bibinfo  {journal} {Reviews of Modern Physics}\ }\textbf
  {\bibinfo {volume} {52}},\ \bibinfo {pages} {299} (\bibinfo {year}
  {1980})}\BibitemShut {NoStop}%
\bibitem [{\citenamefont {Seto}(2025{\natexlab{a}})}]{Seto:2025vud}%
  \BibitemOpen
  \bibfield  {author} {\bibinfo {author} {\bibfnamefont {N.}~\bibnamefont
  {Seto}},\ }\href {https://doi.org/10.1103/PhysRevD.111.083051} {\bibfield
  {journal} {\bibinfo  {journal} {Phys. Rev. D}\ }\textbf {\bibinfo {volume}
  {111}},\ \bibinfo {pages} {083051} (\bibinfo {year} {2025}{\natexlab{a}})},\
  \Eprint {https://arxiv.org/abs/2507.01250} {arXiv:2507.01250 [gr-qc]}
  \BibitemShut {NoStop}%
\bibitem [{\citenamefont {Seto}(2025{\natexlab{b}})}]{Seto:2025vfg}%
  \BibitemOpen
  \bibfield  {author} {\bibinfo {author} {\bibfnamefont {N.}~\bibnamefont
  {Seto}},\ }\href {https://doi.org/10.1103/mqjd-rwrb} {\bibfield  {journal}
  {\bibinfo  {journal} {Phys. Rev. Lett.}\ }\textbf {\bibinfo {volume} {135}},\
  \bibinfo {pages} {061402} (\bibinfo {year} {2025}{\natexlab{b}})},\ \Eprint
  {https://arxiv.org/abs/2506.23441} {arXiv:2506.23441 [astro-ph.HE]}
  \BibitemShut {NoStop}%
\bibitem [{\citenamefont {Jammalamadaka}\ and\ \citenamefont
  {Terdik}(2019)}]{Jammalamadaka2019}%
  \BibitemOpen
  \bibfield  {author} {\bibinfo {author} {\bibfnamefont {S.~R.}\ \bibnamefont
  {Jammalamadaka}}\ and\ \bibinfo {author} {\bibfnamefont {G.~H.}\ \bibnamefont
  {Terdik}},\ }\href {https://doi.org/10.1016/j.jmva.2019.05.007} {\bibfield
  {journal} {\bibinfo  {journal} {Journal of Multivariate Analysis}\ }\textbf
  {\bibinfo {volume} {171}},\ \bibinfo {pages} {436} (\bibinfo {year}
  {2019})}\BibitemShut {NoStop}%
\end{thebibliography}%

\end{document}